\documentclass{mpe_report}

\usepackage{psfig,graphicx,epsfig}
\usepackage{color}
\usepackage{amsmath,amssymb,epic,eepic,array}

\begin{document}

\pagenumbering{arabic}
\setcounter{page}{116}

\renewcommand{\FirstPageOfPaper }{116}\renewcommand{\LastPageOfPaper }{119}%%
%% This is MPE-Report_example.tex
%% LaTeX2e example style file for the contributed talks and posters presented
%% during the 363rd Heraeus Seminar on Neutron Stars and Pulsars, held in 
%% Bad Honnef, May 14.-19. 2006.
%% 
%% This file needs the LaTeX2e class file he_symp.cls  
%%
% -----------------------------------------------------------------------------
%\documentclass{mpe_report}
%\usepackage{psfig}
% -----------------------------------------------------------------------------
%\begin{document}

\title{A multicomponent model for the optical to  $\gamma$-ray emission from the Crab Pulsar}
\author{R. Campana\inst{1} \and E. Massaro\inst{1}, G. Cusumano\inst{2} \and T. Mineo\inst{2}}  
\institute{Department of Physics, University of Rome ``La Sapienza'', Rome, Italy
\and  INAF -- IASF-Pa, Palermo, Italy}
\maketitle

\begin{abstract}
We present a multicomponent model to explain the features of the pulsed emission and spectrum of the Crab Pulsar, on the basis of X and $\gamma$-ray observations performed with BeppoSAX, INTEGRAL and CGRO. This model explains the evolution of the pulse shape and of the phase-resolved spectra, ranging from the optical/UV to the GeV energy band, on the assumption that the observed emission is due to several components. 
The first component, $C_O$, is assumed to have the pulsed double-peaked profile observed at the optical frequencies, while the second component, $C_X$, is dominant in the interpeak and second peak phase regions. The spectra of these components are modelled with log-parabolic laws. 
Moreover, to explain the properties of the pulsed emission in the MeV-GeV band, we introduce two more components, $C_{O\gamma}$ and $C_{X\gamma}$, with phase distributions similar to those of $C_O$ and $C_X$ and log-parabolic spectra with the same curvature but different peak energies. 
This multicomponent model is able to reproduce both the broadband phase-resolved spectral behaviour and the changes of the pulse shape with energy. 
We also propose some possible physical interpretations in which $C_O$ and $C_X$ are emitted by secondary pairs via a synchrotron mechanism while $C_{O\gamma}$ and $C_{X\gamma}$ can originate either from Compton scattered or primary curvature photons. 
\end{abstract}

\section{Introduction}
The study of the phase distributions of pulsars' signals in the various bands of the electromagnetic spectrum is important to obtain information on the 
geometry and location of the emission regions in the magnetosphere.
At $\gamma$-ray energies, in particular, the three brightest sources (Vela, Crab and Geminga)
show remarkably similar patterns with two main peaks at a phase separation ranging from
0.40 to 0.48.

The Crab Pulsar (PSR B0531+21) is  characterized by a rather stable phase distribution throughout the whole electromagnetic spectrum with a double peak structure.
It is well known that the pulse shape of the Crab changes with energy in the X and soft gamma-ray ranges where the emission of the second peak (P2) becomes higher than the first one (P1), and where it is present a significant emission from the region between the two peaks (bridge or interpeak, IP). This behaviour continues up to about 10 MeV, where the pulse almost sharply returns to a shape similar to the optical light curve. A satisfactory explanation for these changes has not been found so far. 

On the basis of high quality BeppoSAX data, covering a wide energy range (from 0.1 to 300 keV), we already proposed a two component model (Massaro \emph{et~al.}, 2000) to explain the behaviour of the light curve.
Here we extend this model, reanalysing the whole set of BeppoSAX Crab observations with new
ISGRI-INTEGRAL data at higher energies (Mineo \emph{et~al.}, 2006). 
We found that the energy spectra of these components are not described by a simple power law, but show a spectral steepening towards high energies. We model these components with log-parabolic spectral distributions. Moreover, to explain the behaviour in the MeV/GeV band as observed by COMPTEL and EGRET onboard Compton-CGRO, two more components are introduced, both with a similar shape and spectrum of the X-ray counterparts.
A complete description of the data analysis and of the model can be found in Massaro \emph{et al.} (2006).
                 
\section{The two-component model: optical to hard X-rays}
  
 As presented in Massaro \emph{et~al.} (2000), Crab X-ray light curve is well reproduced by two phase-components. 
The first component, called $C_O$, is assumed to have the same pulsed profile observed at optical frequencies, while the second component, $C_X$, is described by an analytical model whose shape is determined by comparing $C_O+C_X$ with the observed pulse profiles, and that dominates at the interpeak (IP) and second peak (P2) phase regions (Fig. 1). 

\begin{figure}
\centerline{\psfig{file=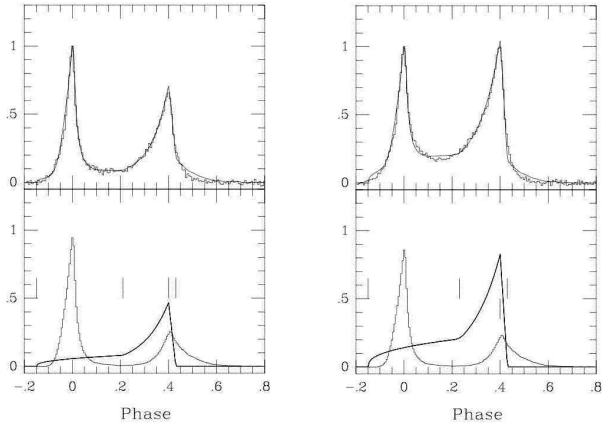,width=8.8cm,clip=} }
\caption{The two components $C_O$ and $C_X$ of the model at the energies of 8.85 keV (left) and 75.2 keV (right). In the upper panels: the model compared with BeppoSAX data. In the lower panels: $C_O$ and $C_X$ (adapted from Massaro \emph{et al.} 2000).
\label{fig1}}
\end{figure}

Using the high-statistics observations of BeppoSAX we performed a phase-resolved spectral analysis and found that the photon indices of P1, P2 and IP are changing with energy, and linearly increasing with Log~E (Fig. 2).
\begin{figure}
\centerline{\psfig{file=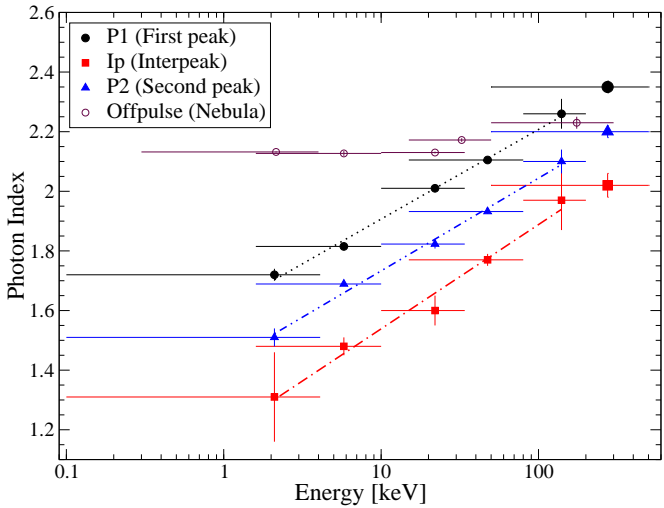,width=7.0cm,angle=-90,clip=} }
\caption{Photon indices of P1, IP and P2 as measured by the four NFI of BeppoSAX and by INTEGRAL-ISGRI.
\label{fig2}}
\end{figure}

We found that the spectra of $C_O$ and $C_X$ are well fitted by a log-parabolic spectral law, 
\begin{equation}
F(E) = KE^{-(a+b\mathrm{Log}E)}
\end{equation}
where $K$ is the flux at 1 keV and $E$ is the energy in keV. The parameter $b$ describes the ``curvature'' of the log-parabola. The energy-dependent spectral index can be obtained from the previous equation: $\alpha(E) = a + 2b~\mathrm{Log}~E$. According to this spectral law, the spectral energy distribution (SED) has a maximum at the energy $E_p = 10^{(2-a)/2b}$. 
The curvature parameter $b$ is equal to 0.16 for both $C_O$ and $C_X$, while the peak energies are respectively 12 keV and 178 keV.

\section{Extension of the model to the MeV/GeV band: the need for two more components}
\begin{figure}
\centerline{\psfig{file=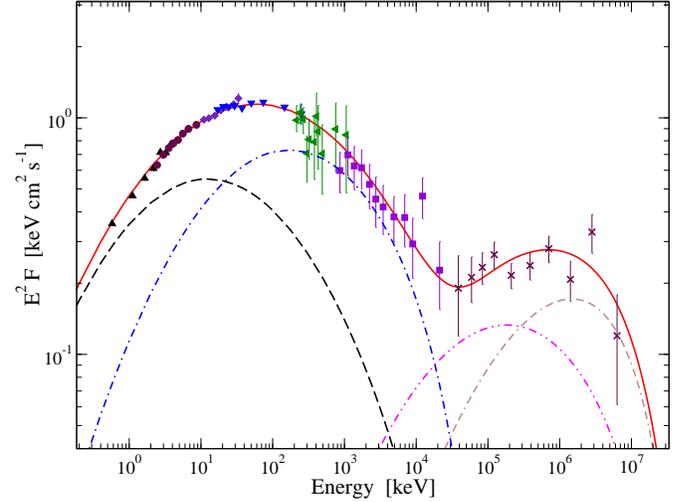,width=7.0cm,angle=-90,clip=} }
\caption{Broadband spectra of the total averaged pulse  with the four components of the model. Upward pointing triangles: LECS; brown circles: MECS; diamonds: HPGSPC; downward pointing triangles: PDS; leftward pointing triangles: FIGARO II; squares: COMPTEL; crosses: EGRET. Dashed line: $C_{O}$; dash-dotted line: $C_{X}$; dot-dot-dashed line: $C_{O\gamma}$; dash-dash-dotted line: $C_{X\gamma}$.
\label{fig3}}
\end{figure}

\begin{figure}
\centerline{\psfig{file=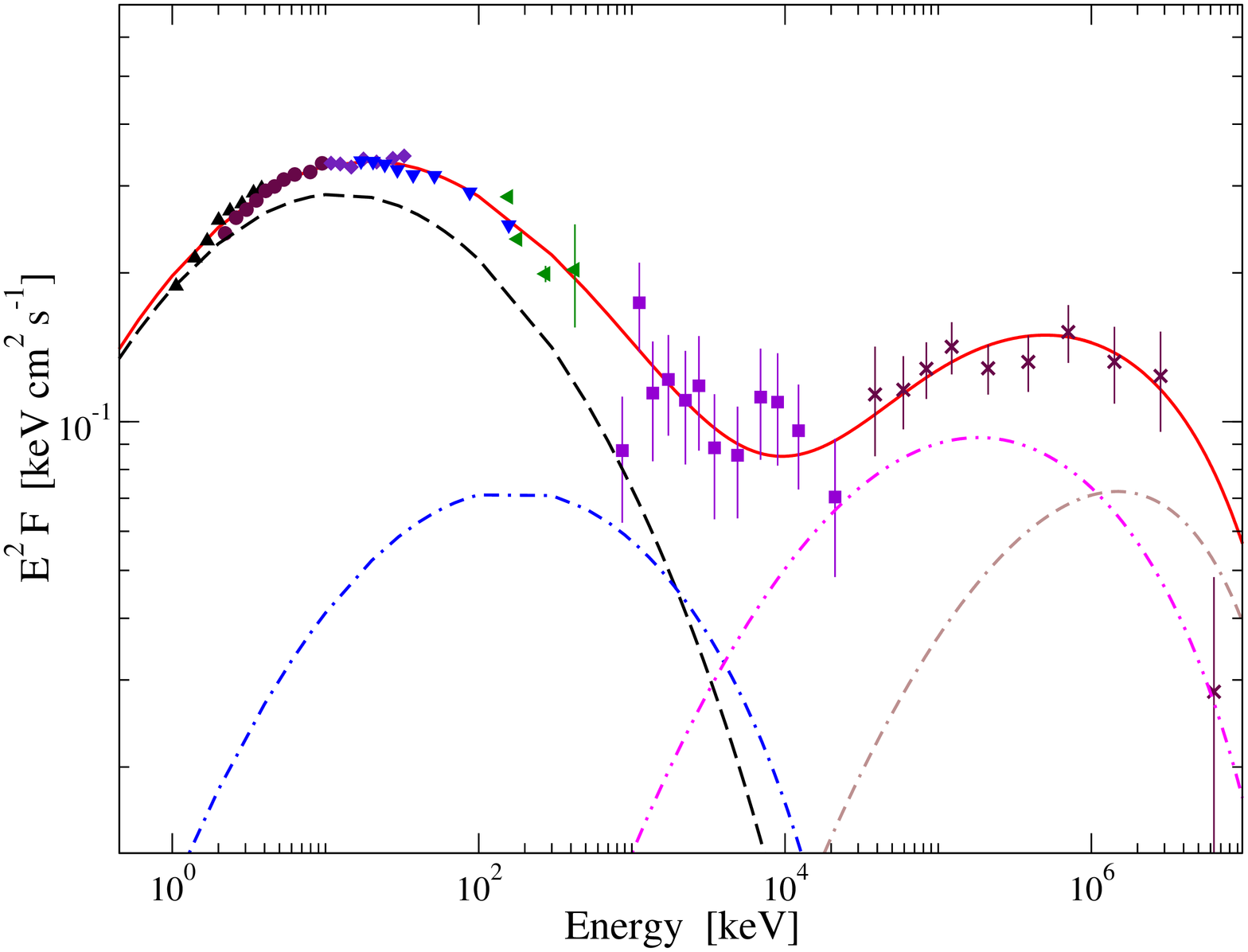,width=7.0cm,angle=-90,clip=} }
\caption{Broadband spectra of P1 with the four components of the model. Upward pointing triangles: LECS; brown circles: MECS; diamonds: HPGSPC; downward pointing triangles: PDS; leftward pointing triangles: ISGRI; squares: COMPTEL; crosses: EGRET. Dashed line: $C_{O}$; dash-dotted line: $C_{X}$; dot-dot-dashed line: $C_{O\gamma}$; dash-dash-dotted line: $C_{X\gamma}$.
\label{fig4}}
\end{figure}

\begin{figure}
\centerline{\psfig{file=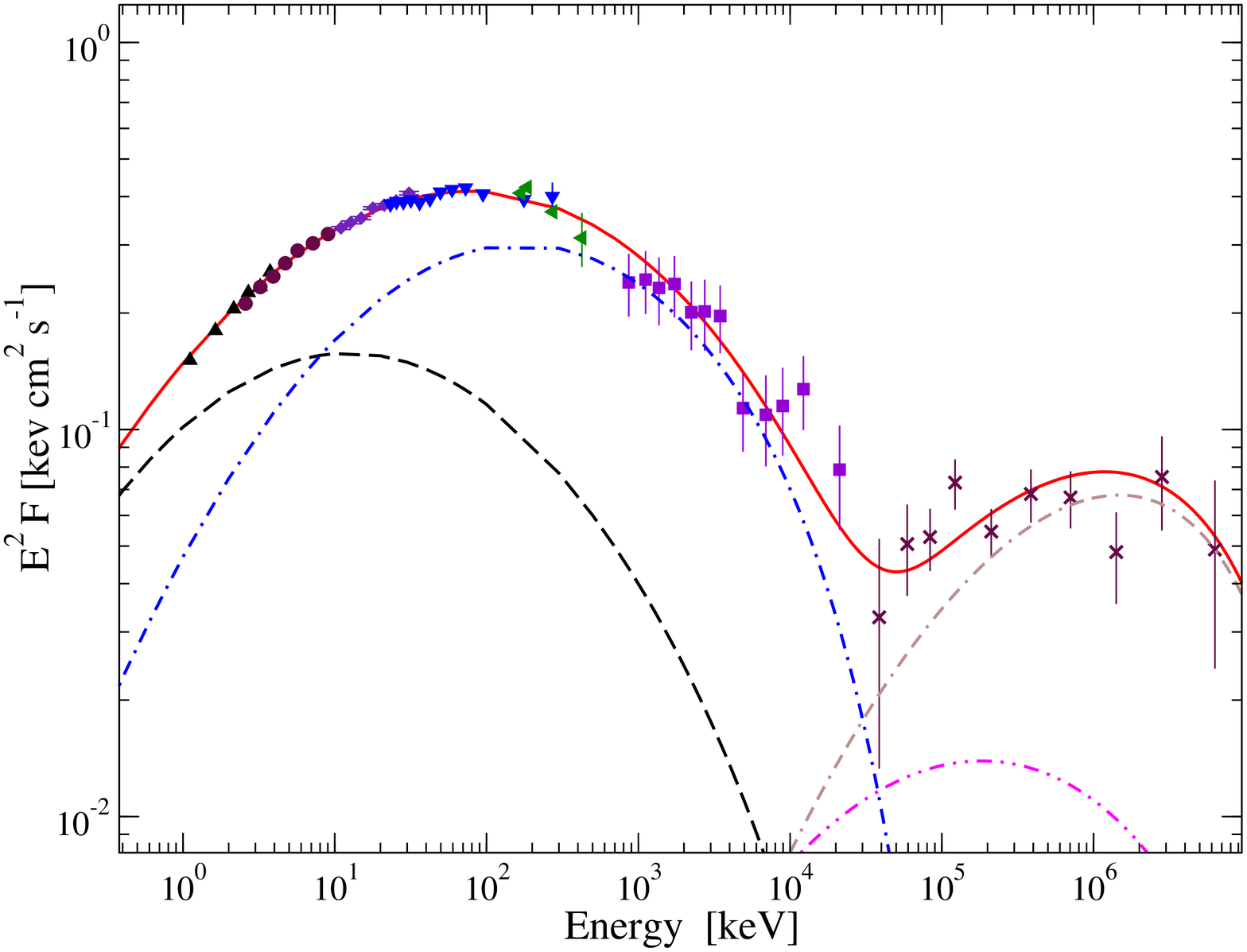,width=7.0cm,angle=-90,clip=} }
\caption{Broadband spectra of P2 with the four components of the model. Upward pointing triangles: LECS; brown circles: MECS; diamonds: HPGSPC; downward pointing triangles: PDS; leftward pointing triangles: ISGRI; squares: COMPTEL; crosses: EGRET. Dashed line: $C_{O}$; dash-dotted line: $C_{X}$; dot-dot-dashed line: $C_{O\gamma}$; dash-dash-dotted line: $C_{X\gamma}$.
\label{fig5}}
\end{figure}

\begin{figure}
\centerline{\psfig{file=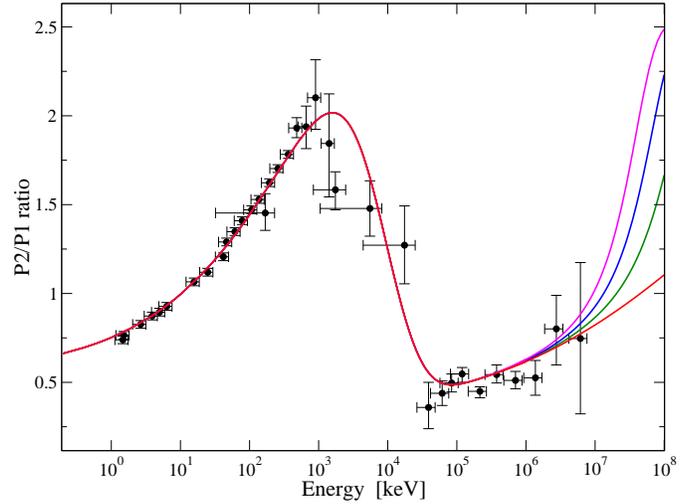,width=7.0cm,angle=-90,clip=} }
\caption{P2/P1 ratio as derived from the model. Data points come from various experiments (Kuiper et al 2001). The various extrapolations above 1 GeV correspond to different values 
of the cut-off energy of the $C_{O\gamma}$ spectrum; from top to bottom: 9, 11,
13 and 15 GeV.
\label{fig6}}
\end{figure}

\begin{figure}
\centerline{\psfig{file=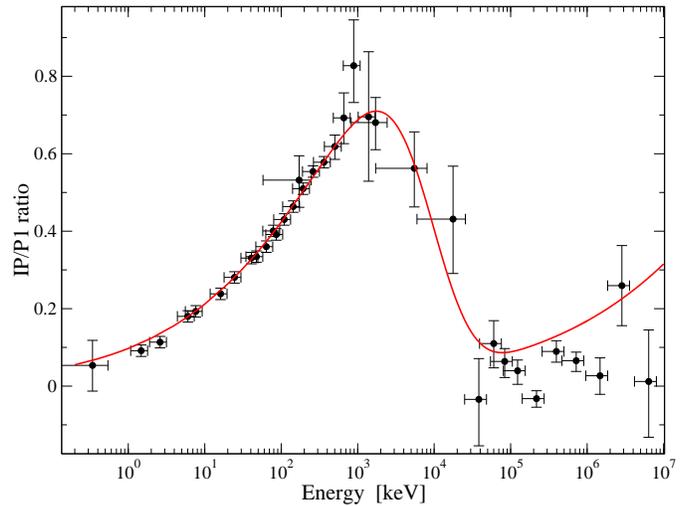,width=7.0cm,angle=-90,clip=} }
\caption{IP/P1 ratio as derived from the model. Data points come from various experiments (Kuiper et al 2001).
\label{fig7}}
\end{figure}

CGRO COMPTEL and EGRET observations (Kuiper \emph{et al.}, 2001; Thompson, 2004) provided above $\sim$10 MeV light curves of a good statistical quality which show that the pulse shape is similar to that of $C_O$, although some minor differences are present. At energies higher than $\sim$500 MeV the emission from IP and P2 increases, and this seems to reproduce the behaviour of the X-ray emission. In order to explain such a finding, we assume that there are two more, high-energy spectral components, $C_{O\gamma}$ and $C_{X\gamma}$, both with a log-parabolic spectral distribution and with the same pulse shape of the lower-energy components $C_O$ and $C_X$. To be consistent with the upper limits to the TeV pulsed emission (e.g. Lessard \emph{et al.}, 2000) we added also an exponential cutoff to both $C_{O\gamma}$ and $C_{X\gamma}$, at the energy $E_c=15$ GeV.
This model therefore has 6 adjustable parameters, i.e. the peak energies, curvatures and normalizations of the $C_{O\gamma}$ and $C_{X\gamma}$ components. Assuming that the curvatures are equal to the $C_O$ and $C_X$ ones ($b = 0.16$),
we are then able to reproduce the broadband energy spectrum of the total (averaged) pulse and of the P1, IP and P2 phase regions (see Figs. \ref{fig3}, \ref{fig4} and \ref{fig5}) and the ratios of P2/P1 and IP/P1 fluxes (in the same phase intervals of Kuiper \emph{et al.}, 2001; Figs. \ref{fig6} and \ref{fig7}). We stress that there is no constraint on $E_c$: in fig. 6 we plot also the P2/P1 ratio for various values of $C_{O\gamma}$ cutoff energy ranging from 9 to 15 GeV.



\section{Physical interpretation}
An open question is the physical origin of these components, that phenomenologically explain the observations with a very good approximation, in the framework of the high-energy pulsar emission models, either in the polar cap or outer gap models (e.g. Cheng \emph{et al.}, 2000; Zhang \& Cheng, 2002). 

Assuming that the lower-energy components $C_O$ and $C_X$ are produced by synchrotron emission of secondary electron-positron pairs created in the pulsar magnetosphere, the higher-energy components $C_{O\gamma}$ and $C_{X\gamma}$ could be due to:
\begin{enumerate}
\item Emission of curvature radiation from primary particles accelerated in the magnetospheric gaps.
\item Emission from inverse Compton scattering of the synchrotron photons by the secondary pairs themselves (Synchrotron-Self-Compton mechanism). 
\end{enumerate}
The different shape of the ``\emph{O}'' and the ``\emph{X}'' components is presumably due to the different location in the magnetosphere of the emission regions.

\section{Conclusions}

Several models have appeared in the literature based on either polar cap or outer gap geometries. Usually, these models are focused on reproducing either the total spectrum or the phase profile, and generally they are not fully satisfactory in explaining the complex observational picture. Moreover, the possibility that the observed features of the pulsed signal can arise from the superposition of two or more distinct components is not taken into account. 

We followed another approach and searched for a possible interpretation 
of the Crab signal based on the superposition of two or more components 
that provides a consistent description of the spectral and phase 
distributions. 
Clearly, it is only a phenomenological model, but it could furnish some constraints to more detailed, physically-based emission models. In particular it is important to verify whether at energies higher than $\sim$1 GeV the pulse shape tends to be dominated by $C_{X\gamma}$. The GLAST/LAT experiment (Gehrels \emph{et al.}, 1999), with its large collecting area, will give us very useful data in this range that will permit to better estimate the model parameters.
 
Another interesting perspective is whether this model can be adapted to the other 
$\gamma$-ray pulsars. 
For Vela and Geminga the main problem is that their pulse profiles change very much in different spectral bands and no clear trend, like the P2/P1 ratio in Crab, has been found
to now. 
In the $\gamma$-ray band, however, Kanbach (1999) showed that the peak ratios of
all these three pulsars have a rather similar behaviour.
This can be an indication that geometrical effects may be more relevant at energies
lower than $\gamma$-rays and that components like $C_O$ or $C_X$, if existing in Vela
and Geminga, are not detected because they are more beamed than the high energy photons.

\begin{acknowledgements}
This work was financially supported by Universit\`a di Roma La Sapienza.
R.C. also acknowledges the support by the WE-Heraeus foundation during the seminar.
\end{acknowledgements}

%\end{document}
          \clearpage


\begin{thebibliography}{9}
	\bibitem{cheng} {Cheng K. S., Ruderman M., and Zhang L.},
	\textsl{ApJ}, \textbf{537}, 964-976, (2000).
	\bibitem{gehrels} {Gehrels N., Michelson P., \emph{et al.}},
	\textsl{Aph}, \textbf{11}, 277-282, (1999).
	\bibitem{kanbach} {Kanbach G.}, 
	Proc. 3rd INTEGRAL Workshop, \textsl{Astrophys. Lett. Comm.}, 
	\textbf{38}, 17, (1999).
	\bibitem{kuiper} {Kuiper L., Hermsen W., Cusumano G., \emph{et al.}}, 
	\textsl{A\&A}, \textbf{378}, 918-935, (2001).
	\bibitem{lessard} {Lessard R. W., Bond I. H., Bradbury S. M., \emph{et al.}},
	\textsl{ApJ}, \textbf{531}, 942-948, (2000).
	\bibitem{massaro2000} {Massaro E., Cusumano G., Litterio M., and Mineo T.},
	\textsl{A\&A}, \textbf{375}, 397-404, (2000).
         \bibitem{massaro2006} {Massaro E., Campana R., Cusumano G., and Mineo T.},
	\textsl{A\&A}, \textbf{459}, 859-870, (2006).
	[astro-ph/0607410]
	\bibitem{mineo2006} {Mineo T., Ferrigno C., Foschini L. \emph{et al.}},
	\textsl{A\&A}, \textbf{450}, 617-623, (2006).
	\bibitem{thompson} {Thompson D. J.}, in
	\textsl{Cosmic Gamma Ray Sources}, Kluwer ASSL series, \textbf{304}, edited by K. Cheng and G. Romero, (2004). [astro-ph/0312272]
	\bibitem{zhang} {Zhang L. and Cheng K. S.},
	\textsl{ApJ}, \textbf{569}, 872-877, (2002).
 \end{thebibliography}
\end{document}